\documentclass[onecolumn,floatfix,nofootinbib]{revtex4-1}
\usepackage[english]{babel}

\usepackage{amsthm}
\usepackage{amsmath}
\usepackage{latexsym}
\usepackage{amsfonts}
\usepackage{amssymb}
\usepackage{epstopdf}

\usepackage[dvipsnames]{xcolor}
\usepackage{bbm,dsfont}
\usepackage{graphicx}
\usepackage{mathrsfs}
\usepackage{mathbbol}
\usepackage{hyperref}
\usepackage{enumerate}
\usepackage{MnSymbol}
\usepackage{bbding}
\usepackage{float}
\usepackage{soul}
\usepackage{ulem}
\usepackage{physics}

\begin{document}

\title{Metastable quantum entrainment}

\author{Albert Cabot}
\affiliation{IFISC (UIB-CSIC), Instituto de F\'isica Interdisciplinar y Sistemas Complejos, Palma de Mallorca, Spain}

\author{Gian Luca Giorgi}
\affiliation{IFISC (UIB-CSIC), Instituto de F\'isica Interdisciplinar y Sistemas Complejos, Palma de Mallorca, Spain}

\author{Roberta Zambrini}
\affiliation{IFISC (UIB-CSIC), Instituto de F\'isica Interdisciplinar y Sistemas Complejos, Palma de Mallorca, Spain}

\begin{abstract}
Quantum van der Pol oscillators are driven-dissipative systems displaying quantum synchronization phenomena. When forced by a squeezed drive, the  frequency adjusts to half of the forcing displaying multiple preferred phases. Here we analyze the physical origin of this entrained response, establishing a connection with metastability in open quantum systems. We report a dynamical regime characterized by a huge separation of time scales, in which a dynamical mode displays a lifetime that can be orders of magnitude larger than the rest. In this regime, the long-time dynamics is captured by an incoherent process between two metastable states, which correspond to the preferred phases of the synchronized oscillator. In fact, we show that quantum entrainment  is here characterized by fluctuations driving an incoherent process between two metastable phases, which ultimately limits its temporal coherence when moving into the quantum regime. Finally, we discuss connections with the phenomena of dissipative phase transitions and transient synchronization in open quantum systems.
\end{abstract}

\maketitle

\section{Introduction}
\label{sec:intro}

The investigation of the properties of the nonequilibrium dynamics of dissipative quantum systems has  recently become a major research subject,  with focus ranging from their dynamics and control, to phase transitions, collective phenomena or thermodynamic features \cite{WisemanMilburn,DPT_exp1,DPT_exp2,carusotto,hartmann,nissen,diehl,Talkner2020}.
One of the consequences of the interplay between driving, dissipation and interactions is the possibility of having the dynamics of an open quantum system dominated  by a huge separation of time scales, a characteristic of metastability \cite{Garrahan1}. In this case, initially highly excited configurations quickly relax to a set of metastable states that act as attractors of the dynamics for a long intermediate timescale, until final relaxation to the true stationary state occurs. This multi-step dynamics  has different physical origins, as population trapping in a driven three-level system \cite{Garrahan1}, the coexistence of different phases in an open Ising model \cite{Garrahan2} or in a non-linear oscillator \cite{Armour2020}, a dissipative discrete-time crystal \cite{Gambetta}, the long-time response of a driven-dissipative Rabi model \cite{Plenio2}, or a quasistationary regime in a chain of interacting fermions subject to a localized particle loss \cite{Wolff}. A general approach towards metastability in open quantum systems has been recently introduced in Ref. \cite{Garrahan1}, in which some of its crucial signatures in the spectral properties of the Liouvillian have been identified.

The characteristic timescale separation of metastability has also been found in between different dynamical regimes of driven-dissipative systems \cite{Vogel,DPT_Casteels2}, or as  a precursor of eventual spontaneous symmetry breaking in the thermodynamic or infinite-excitation limits of these systems \cite{Minganti2018}. Indeed, sudden changes can occur in the stationary state of these systems in these limits, what are known as dissipative phase transitions (DPTs) \cite{Kessler,Minganti2018}. As in the case of quantum phase transitions, DPTs are associated with a  closure of the spectral gap of the generator of the dynamics \cite{Kessler,Minganti2018}. This asymptotic gap closure has profound consequences even far from the thermodynamic or classical limits, as it can lead to a very slow relaxation timescale and hence metastability in the quantum dynamics \cite{Garrahan1}. While for first order DPTs this usually occurs in a narrow region between the different regimes \cite{DPT_Casteels2,DPT_Casteels1}, as experimentally observed in Refs. \cite{DPT_exp1,DPT_exp2}, for symmetry-breaking DPTs this occurs in a whole dynamical regime characterized by the emergence of metastable symmetry-broken states \cite{Minganti2018,Minganti2016,Bartolo2016b}.

A paradigmatic phenomenon of classical driven-dissipative systems is synchronization, which emerges in widely different contexts and forms \cite{Pikovsky,Belanov}. Entrainment is a type of synchronization phenomenon in which a self-oscillating system \cite{Strogatz} adjusts its frequency and phase to that of an external forcing or to a multiple/fraction of it. Otherwise, synchronization can also occur as a mutual or spontaneous phenomenon among interacting system components. In the past decade, intense research activity has addressed the fundamental question of whether synchronization can emerge also in quantum systems \cite{SyncRev1}. As for its classical counterpart, diverse approaches and systems have been considered, since the seminal works of Refs. \cite{Hanggi,Zhirov}. Different signatures and  indicators of quantum synchronization have been explored in the literature \cite{SyncRev1,Giorgi1,Lee2013,Bruder2014,Armour2015,Mari,Roulet_PRA,generalmeasure,Bruder2016}, while its emergence has been reported  in disparate scenarios as systems of spins \cite{Giorgi2,Bruder2018,Armour2015}, networks of quantum harmonic oscillators \cite{Manzano_SciReps,Cabot_NPJ}, optomechanical systems \cite{Marquardt,Weiss2},  atomic lattices \cite{Cabot_PRL,Dieter2} and clouds \cite{Holland1,Holland2}, and micromasers \cite{Armour2018}. Moreover, novel forms of synchronization have been identified \cite{Bruder2016,Dieter2,Lee2014,buca2021}, as the case of transient synchronization \cite{SyncRev2}, in which spontaneous synchronization emerges due to a long-lived collective excitation of the quantum system, a phenomenon that has been linked to the presence of super/subradiance \cite{Bellomo} and long-lived correlations \cite{Giorgi1,Giorgi2,Manzano_SciReps,Cabot_NPJ,Manzano_PRA,Olaya}.

A rich playground where  synchronization can be extended from the classical to the quantum regime is offered by  the  paradigmatic van der Pol oscillator (vdP) \cite{Pikovsky,Belanov}, introduced in  Refs. \cite{Lee2013,Bruder2014}. For this model, both entrainment by a harmonic drive and spontaneous synchronization between coupled oscillators have been found in the quantum regime \cite{Lee2013,Bruder2014,Lee2014,Bruder2015}. 
Signatures of synchronization have been identified in phase-space representations of the stationary state as well as in the power spectrum. Indeed, the stationary Wigner distribution \cite{Carmichael} of the quantum van der Pol  (QvdP) displays a ring-like shape in the absence of entrainment, which turns to a localized lobe for large enough driving strength \cite{Lee2013,Bruder2014}. Moreover, the frequency of the maximum of the power spectrum is reported to shift towards that of the driving as the system becomes entrained \cite{Bruder2014}. An important caveat is that the more in the quantum regime the system operates, the harder it  is to synchronize it due to the detrimental effect of quantum fluctuations \cite{Bruder2014}. As an alternative implementation strategy, a squeezed forcing has been reported to enhance entrainment in the quantum regime \cite{Sonar2018}. However, very deep into it, where the system behaves effectively as a few-level system, the enhancing effect of squeezing is limited \cite{Sq_PRR}.

The introduction of the squeezed forcing comes at the expense of changing qualitatively the dynamical scenario of entrainment \cite{Sonar2018,Kato2019,Kato2021}. This is an observation of utmost importance to understand the entrained quantum dynamics of this system, as we show here. While in the classical harmonically driven vdP oscillator entrainment corresponds to a fixed point attractor  (in the rotating frame) \cite{Pikovsky,Weiss}, here it corresponds to bistability (in the rotating frame) \cite{Sonar2018,Kato2021}, characterized by frequency locking to half of the frequency of the forcing and  two possible  locked-phases. As it turns out, the nature of the classical attractors has a deep influence in the quantum  dynamics. In fact, for the driven QvdP, the entrained dynamics  has been shown to be well described by a linearized model around the fixed point attractor \cite{Weiss}. This approach sheds light on the synchronized dynamics of this system and on the reported behavior of the power spectrum, while it unveils new dynamical features as phase-coherent dynamics \cite{Weiss}.  In contrast, we show that the squeezed QvdP displays a distinct dynamical scenario in which the main features of the entrained response cannot be captured by a linearized model, a fact rooted in the emergence of bistability in the classical limit.

Motivated by the rich physics behind the squeezed vdP oscillator, we show in this work  that such a model offers  the opportunity to
 establish a connection between metastability  and  entrainment, that we argue to transcend this specific system and to be applicable to other synchronization scenarios in the  presence of phase multistability. 
More specifically, we identify the huge timescale separation characteristic of a metastable dynamics as  the distinctive feature of the entrained dynamics for this system. 
On the shortest time scale, any initial state rapidly relaxes to the manifold of states spanned by two metastable states. These are the two preferred phases of the oscillator when it is well entrained by the forcing. On the longest time scale, the dynamics is dominated by a slow fluctuation mode connecting both phases, which eventually drives the state of the system to an even incoherent mixture of both. In between, there is a long period of time in which the response of the system is a quasi-coherent subharmonic oscillation.  This is precisely how subharmonic entrainment manifests in the quantum regime, and it dominates the behavior of the power spectrum. While these phases act as effective attractors of the dynamics (any initial condition quickly relaxes to them), quantum fluctuations  turn phase bistability to phase metastability, thus limiting the temporal coherence of the synchronized response on the long-time.

\section{The Model}\label{SecModel}

In the laboratory frame, the squeezed van der Pol oscillator is described by the time-dependent master equation for the state of the system $\hat{\rho}_l$ ($\hbar=1$) \cite{Sonar2018}:
\begin{equation}\label{ME}
\begin{split}
\partial_t\hat{\rho}_L=-i[\hat{H}_L(t),\hat{\rho}_L]+\frac{\gamma_1}{2}\mathcal{D}[\hat{a}^\dagger]\hat{\rho}_L+\frac{\gamma_2}{2}\mathcal{D}[\hat{a}^2]\hat{\rho}_L,\\
\hat{H}_L(t)=\omega_0 \hat{a}^\dagger \hat{a}+i\eta(\hat{a}^2e^{i2\omega_s t}-\hat{a}^{\dagger2}e^{-i2\omega_s t}),
\end{split}
\end{equation}
where we have defined the dissipator $\mathcal{D}[\hat{L}]\hat{\rho}=2\hat{L}\hat{\rho}\hat{L}^\dagger-\hat{L}^\dagger \hat{L}\hat{\rho}-\hat{\rho}\hat{L}^\dagger \hat{L}$, $\hat{L}$ is the corresponding jump operator, $\hat{a}$ is the annihilation operator of the bosonic mode,  $\gamma_1$ is the amplification rate, $\gamma_2$ the nonlinear damping rate, $\eta$ is the squeezing strength of frequency $2\omega_s$, and $\omega_0$ is the intrinsic frequency of the system. The explicit time dependence of this model enters only through the Hamiltonian term, which can be eliminated in a rotating frame, as can be achieved by a standard time-dependent unitary transformation, $\hat{U}_t=\text{exp}(-i\omega_s\hat{a}^\dagger \hat{a}t)$. This leads to a master equation with the same dissipative part but with a time-independent Hamiltonian\footnote{The rotating frame and laboratory frame Hamiltonians are related by $\hat{H}=\hat{U}_t^\dagger \hat{H}_L(t)\hat{U}_t-i\hat{U}_t^\dagger\partial_t \hat{U}_t$, while the state of the system transforms as $\hat{\rho}(t)=\hat{U}_t^\dagger \hat{\rho}_L(t)\hat{U}_t$.}: $\hat{H}=\Delta \hat{a}^\dagger \hat{a}+i\eta(\hat{a}^2-\hat{a}^{\dagger2})$ with $\Delta=\omega_0-\omega_s$. Notice that we denote the state of the system in this  frame as $\hat{\rho}$, without any subscript, while the laboratory frame is indicated by the subscript '$L$'.

Possible experimental implementations of the QvdP oscillator have been described in the literature, considering platforms of trapped ions \cite{Lee2013,Sonar2018} and optomechanical oscillators \cite{Bruder2014,Bruder2015,Sonar2018}. In both cases, the QvdP oscillator corresponds to  the properly engineered effective dynamics of a mechanical degree of freedom. Then, amplification and non-linear dissipation can be implemented by driving the fast optical degrees of freedom with lasers resonant with  one-phonon absorption and two-phonon emission processes. Moreover, several possibilities for the implementation of the squeezed (two-boson) drive of Eq. (1) have been presented in \cite{Sonar2018} too, considering both optomechanical systems (e.g. modulating electrically the spring constant of the mechanical mode \cite{Rugar1991})  and trapped ions (e.g.  by a combination of two Raman beams detuned by $2\omega_s$ \cite{Meekhof1996}).

In this rotating frame,  the master equation can be written in terms of the $time$-$independent$ Liouvillian superoperator, defined by $\partial_t\hat{\rho}=\mathcal{L}\hat{\rho}$. Then, the dynamics of the system can be analyzed considering the eigenspectrum of this superoperator, which is composed by a set of eigenvalues $\lambda_j$, and the corresponding set of right and left eigenmatrices defined by $\mathcal{L}\hat{\rho}_j=\lambda_j\hat{\rho}_j$ and  $\mathcal{L}^\dagger\hat{\sigma}_j=\lambda^*_j\hat{\sigma}_j$, respectively \cite{Minganti2018,VAlbert} (more details in \ref{Sec_Liouvillian}). Generally, we can use the Liouvillian eigenmodes to decompose the state of the system at any time
\begin{equation}\label{general_dynamics}
\hat{\rho}(t)=\hat{\rho}_{ss}+\sum_{j\geq 1}\text{Tr}[\hat{\sigma}_j^\dagger\hat{\rho}(0)]\hat{\rho}_je^{\lambda_j t}. 
\end{equation}
Similarly, the eigenmode decomposition gives insight in the dynamics of expected values as well as multitime correlations, as we show below. In this expression we have defined the stationary state as $\hat{\rho}_{ss}=\hat{\rho}_{0}/\text{Tr}[\hat{\rho}_0]$, being the corresponding zero eigenvalue $\lambda_0=0$. Notice that the real part of these eigenvalues are non-positive, and we order them such that $\text{Re}[\lambda_0]\geq\text{Re}[\lambda_1]\geq \text{Re}[\lambda_2]\geq \dots$.
In the following, we will refer to the decay rates and frequencies of the eigenmodes
\begin{equation}\label{GAmmaNu}
\Gamma_j=|\text{Re}[\lambda_j]| ~ \text{ and } ~ \nu_j=\text{Im}[\lambda_j], 
\end{equation}
respectively.

While the formalism is fully quantum, we can identify the imbalance between  nonlinear dissipation and linear amplification $\gamma_2/\gamma_1$ as the physical quantity controlling the excitation strength (i.e. boson number) in the family of QvdP systems \cite{Lee2013,Bruder2014}, and hence the classical versus quantum system operation. In the limit $\gamma_2/\gamma_1\gg1$ the QvdP oscillator is well approximated by a two-level system \cite{Bruder2014,Lee2014,Sq_PRR}, while in the limit $\gamma_2/\gamma_1\ll1$ the typical number of excitations becomes macroscopic. Indeed, in Ref. \cite{us} we analyze in detail how the classical attractors emerge in this model as $\gamma_2/\gamma_1\to0$. Here instead, we will focus on intermediate values of $\gamma_2/\gamma_1$ for which entrainment in the quantum regime has been reported \cite{Sonar2018}.

A brief overview of the classical dynamics is presented in \ref{Sec_class}, in which we show the system to display two different dynamical regimes depending solely on the relation between squeezing and detuning, $\eta$ and $\Delta$. In the rotating frame,  for $\eta<\eta_c$ the stable attractor is a limit-cycle, while for $\eta>\eta_c$ the stable attractors are two fixed points. At the classical critical point $\eta_c=|\Delta|/2$  the  bifurcation between the two regimes occurs. In the laboratory frame the limit-cycle regime corresponds to the lack of entrainment, as generally its frequency is not commensurate with that of the forcing. On the other hand, the bistable fixed points oscillate harmonically at half of the frequency of the forcing, and thus this regime corresponds to subharmonic entrainment. Then, each of the fixed points corresponds to one of the possible bistable locked-phases, with a characteristic difference of phase of $\pi$.

\section{timescale separation and metastability}\label{sec_meta_Liouvillian}

In this section we report on the signatures of metastability in the framework of the  Liouvillian analysis: the presence of an eigenmode with a lifetime much longer than the rest, i.e. the opening of a spectral gap, and how this can be exploited to derive a simplified effective picture for the long-time dynamics of the system.

\subsection{Opening of a spectral gap}\label{sec_gap}

\begin{figure}[t!]
\includegraphics[width=0.75\linewidth]{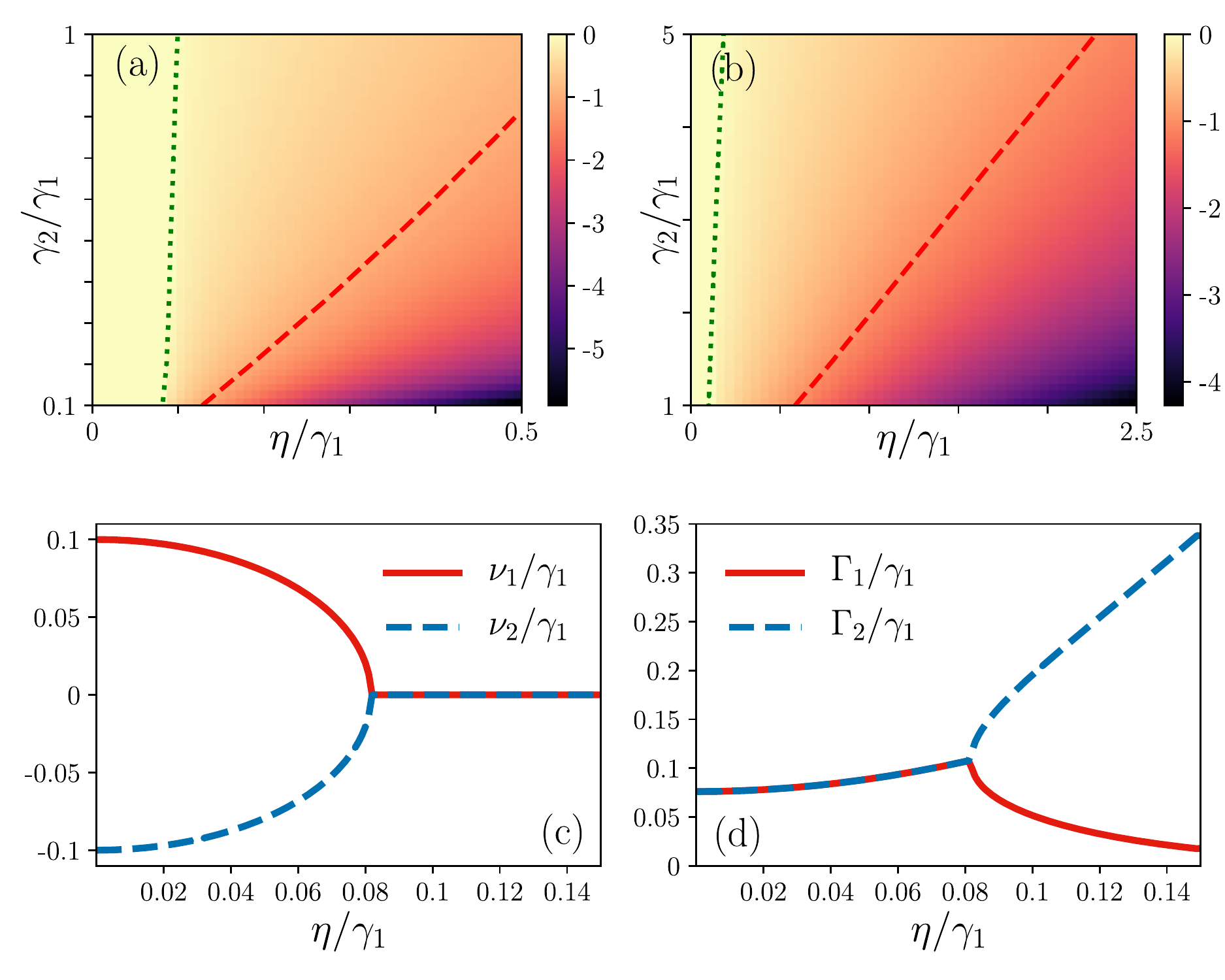}\\
\caption{Colormaps: $\text{log}_{10}(\Gamma_1/\Gamma_2)$ varying $\gamma_2/\gamma_1$ and $\eta/\gamma_1$. Panel (a) covers in detail the small values of $\gamma_2/\gamma_1$ and $\eta/\gamma_1$, while panel (b) displays a wider range of larger values. Green-dotted lines: contour limiting the region in which $\Gamma_1/\Gamma_2=1$. Red-dashed lines: contours indicating $\Gamma_1/\Gamma_2=0.1$. (c) and (d), eigenfrequencies and decay rates of the first two eigenmodes varying $\eta/\gamma_1$, respectively, and for $\gamma_2/\gamma_1=0.1$. In all panels $\Delta/\gamma_1=0.1$.  Notice that in all this work the numerical results  are obtained after truncation of the Fock space to a large enough Fock state such that the convergence of the results with this truncation can be assured. The needed truncation size generally increases as $\gamma_2/\gamma_1$ is diminished and as $\eta/\gamma_1$ is increased, and for the parameter regimes explored in this work this is bounded to the first 50 levels. }\label{F1}
\end{figure}

The most important characteristic of the Liouvillian  spectrum of the squeezed QvdP oscillator is that there is a whole parameter region where the minimum decay rate $\Gamma_1$ can be orders of magnitude smaller than $\Gamma_{j\geq2}$. The inverse of such a small decay rate  determines the longest relaxation timescale of our system.
In Fig. \ref{F1} we demonstrate and analyze the formation of such a gap between the decay rates, while in the forthcoming sections we discuss its dynamical consequences and its relation to quantum entrainment.

The spectral gap is characterized through the ratio $\Gamma_1/\Gamma_2$ and can vary several orders of magnitude depending on the squeezing strength and the non-linear damping, Fig. \ref{F1} (a) and (b). As commented, we focus on parameter regimes for which the QvdP oscillator displays a moderate boson number, being neither strongly confined to the two lowest levels nor in the classical limit of large population. As the excitation number grows not only with the linear amplification but also with the squeezing strength, this corresponds to intermediate values of $\gamma_2/\gamma_1$ and small squeezing  $\eta/\gamma_1$ (Fig. \ref{F1} (a)) or to larger non-linear damping while increasing the squeezing  (Fig. \ref{F1} (b)).

The ratio of decay rates displays the following characteristic dependence on the squeezing strength $\eta$: below a certain threshold value, this ratio remains constant and equal to one (bright region), while further increasing  $\eta$ the ratio diminishes steeply (dark region). This threshold value corresponds to an exceptional point (EP) of the Liouvillian \cite{heiss,Minganti2019}, at which $\lambda_{1,2}$ become the same (Fig. \ref{F1} (c-d)) while the corresponding eigenmatrices coalesce (not shown). The green-dotted line in Fig. \ref{F1} (a-b)
indicates the squeezing strength  at which the EP occurs  ($\eta_{EP}$)  as $\gamma_2/\gamma_1$ is varied. For $\eta<\eta_{EP}$, $\lambda_{1,2}$ are complex conjugates, hence explaining the fact that $\Gamma_1/\Gamma_2=1$ for this region. While for $\eta>\eta_{EP}$, $\lambda_{1,2}$ become real valued and $\Gamma_{1}$ decreases when increasing the squeezing strength, while $\Gamma_2$ increases, as exemplified in Fig. \ref{F1} (d). 

Although the behavior of $\Gamma_1/\Gamma_2$ is qualitatively the same for all the considered range of $\gamma_2/\gamma_1$, we find that there are important quantitative differences. In particular, the smaller is $\gamma_2/\gamma_1$ the wider is the gap for a given squeezing strength. This  is intimately related to the behavior of $\Gamma_1$ as the classical limit is approached (i.e. $\gamma_2/\gamma_1\to0$). Indeed, in this limit it is found that $\Gamma_1\to0$ for $\eta>\eta_c$ while $\Gamma_2$ saturates to a non-zero value \cite{us}. This means that the steady state becomes degenerate and the system undergoes a DPT \cite{Kessler,Minganti2018}. This is intimately related to the emergence of the classical bistable attractors as we explain in detail in \cite{us}. Notice that in this limit it is also found that $\eta_{EP}\to\eta_c$.

In broad terms,  the spectral gap is enhanced in presence of a large number of excitations: the larger is the non-linear damping, the more squeezing is needed. The dynamics of the system  displays a significant timescales separation for rates $\Gamma_1$ and $\Gamma_2$ differing by an order of magnitude or more,  as in the (right) regions delimited by a red-dashed line  (where $\Gamma_1/\Gamma_2\leq 0.1$) in Fig. \ref{F1} (a) and (b).

\subsection{Effective long-time dynamics}

The disparity between the weakest damping rate $\Gamma_1$ and the  others allows to capture the system dynamics after a transient considering only  the disparate timescales $\tau_{1,2}^{-1}=\Gamma_{1,2}$. This large separation of time scales makes an initially excited state to rapidly decay  ($\sim\tau_2$) to the manifold spanned by $\hat{\rho}_{ss}$ and the longest-lived eigenmode $\hat{\rho}_{1}$, where it displays a slow relaxation to $\hat{\rho}_{ss}$ at times of the order of $\tau_1$ \cite{Garrahan1,Garrahan2}.  In the presence of a significant spectral gap, as reported in the previous section, there is a well-defined intermediate time scale, $\tau_2\ll t\ll \tau_1$, in which the contribution of the higher modes is negligible while the decay of $\hat{\rho}_1$ is not yet appreciable, and thus the state of the system
appears stationary  in this time window. This is actually a signature of {\it metastability} \cite{Garrahan1}, as we develop further in this section.

After times of the order of $\tau_2$, the dynamics is well-approximated just considering the contributions of the longest-lived eigenmode and of the stationary state. As we have shown in the previous section, the parameter regime in which  $\Gamma_1/\Gamma_2\ll1$ corresponds to $\eta>\eta_{EP}$, and thus $\lambda_{1,2}$ are real valued while the corresponding right and left eigenmatrices are Hermitian \cite{Minganti2018,VAlbert}. Then the long-time dynamics can be approximated as
\begin{equation}\label{MMdynamics}
\hat{\rho}(t)\approx\hat{\rho}_{ss}+\text{Tr}[\hat{\sigma}_1\hat{\rho}(0)]\hat{\rho}_1e^{-\Gamma_1 t}, 
\end{equation}
which follows from Eq. (\ref{general_dynamics}) neglecting  the contributions for the modes with $j\geq2$.  By using the formalism introduced in Refs. \cite{Garrahan1,Garrahan2}, the approximate long-time  dynamics of Eq. (\ref{MMdynamics}) can be recasted in terms of an effective stochastic process between two particular {\it metastable states}, which provides an insightful representation of this long-time response in a basis different from that provided by the stationary state and the longest-lived eigenmode $\hat{\rho}_1$, the latter not being a physical state since it is traceless \cite{Minganti2018}. In the following, we discuss the application of such formalism \cite{Garrahan1,Garrahan2} to our system (further details are in  \ref{Sec_meta_states}).

From the long-time approximation of Eq. (\ref{MMdynamics}) it follows that the state of the system is restricted to the convex manifold of states  spanned by the projection of the initial condition over the stationary state and the longest-lived eigenmode:
\begin{equation}
\mathcal{P}\hat{\rho}=\hat{\rho}_{ss}+\text{Tr}[\hat{\sigma}_1\hat{\rho}]\hat{\rho}_1.
\end{equation}
We denote this manifold as the metastable manifold, as it captures the state of the system for the intermediate time scale $\tau_2\ll t\ll\tau_1$. In Refs. \cite{Garrahan1,Garrahan2}, it is shown that the metastable manifold can be parametrized in terms of the extreme metastable states (EMSs), defined as:
\begin{equation}\label{EMSs}
\hat{\mu}_1 =\hat{\rho}_{ss}+c_{max}\hat{\rho}_1,\quad \hat{\mu}_2=\hat{\rho}_{ss}+c_{min}\hat{\rho}_1,
\end{equation}
where $c_{max}$ and $c_{min}$ are the maximum and minimum eigenvalues of $\hat{\sigma}_1$. In our system, numerical analysis reveals that $c_{min}=-c_{max}$  for the considered regime, i.e. when $\Gamma_1/\Gamma_2<1$. Moreover, we find that for a significant spectral gap ($\Gamma_1/\Gamma_2\ll1$) these coefficients can be well approximated by $c_{max}=-c_{min}\approx1$, which yields the simpler expressions for the EMSs: $\hat{\mu}_{1(2)}\approx \hat{\rho}_{ss}+(-) \hat{\rho}_1$ (see  \ref{Sec_meta_states}). The projection of the state of the system at a given time onto the metastable manifold can be written in terms of the EMSs as:
\begin{equation}\label{entrainment_EMSs2}
\mathcal{P}\hat{\rho}(t)=p_1(t)\hat{\mu}_1+p_2(t)\hat{\mu}_2
\end{equation}
where the real coefficients $p_{1,2}(t)$ are defined in appendix \ref{Sec_meta_states}, while they are constraint  to satisfy \cite{Garrahan1,Garrahan2}:
\begin{equation}\label{entrainment_Probabilities}
p_{1,2}(t)\geq0,\quad p_1(t)+p_2(t)=1.
\end{equation}
By means of the EMSs the long-time dynamics can be parametrized in terms of two physical states\footnote{Notice that $\hat{\mu}_{1,2}$ are Hermitian with trace one, but only approximately positive. The small corrections to positivity arise after neglecting the contribution of the higher modes in Eq. (\ref{MMdynamics}). Thus, the smaller is $\Gamma_1/\Gamma_2$, the better is the approximation to neglect them for $t\gg\tau_2$ and, accordingly, the smaller are these corrections \cite{Garrahan1}.}, whose properties can be readily characterized. Crucially, the EMSs constitute the only basis for which  $p_{1,2}(t)$ satisfy Eq. (\ref{entrainment_Probabilities}) in the  whole metastable manifold. Indeed, in this case $p_{1,2}(t)$ can be interpreted as probabilities and the dynamics within the metastable manifold is given by \cite{Garrahan1,Garrahan2}: 
\begin{equation}\label{eff_dyn}
\dot{p}_1(t)=-\dot{p}_2(t)=-\frac{\Gamma_1}{2}\big[p_1(t)-p_2(t)\big],
\end{equation}
whose solution makes Eq. (\ref{entrainment_EMSs2})  equivalent to  Eq. (\ref{MMdynamics}). Thus, the approximate long-time dynamics for the state of the system has been recasted in the form of a two-state stochastic process between the EMSs,  with a switching rate $\Gamma_1/2$ \cite{Gardiner}. This characterizes the slow dynamics to which an initially excited state rapidly relaxes.  From the solution of this dynamics (\ref{Sec_meta_states}) we find that $p_1(t\to \infty)=p_2(t\to\infty)=1/2$, which could be also deduced from the fact that $\hat{\rho}_{ss}=(\hat{\mu}_1+\hat{\mu}_2)/2$. This illustrates the notion of metastability in our system: any initial unbalanced mixture of the EMSs decays on a very long-time scale to the balanced one ($\hat{\rho}_{ss}$), remaining apparently stable for the well-defined intermediate time scale $\tau_2\ll t\ll\tau_1$. In the following section we explore in detail the physical meaning of this statement by characterizing  the properties of $\hat{\mu}_{1,2}$, through a visual representation based on the Wigner function, and of this effective incoherent dynamics between them.

\section{Metastable preferred phases}

\begin{figure}[t!]
\includegraphics[width=0.8\linewidth]{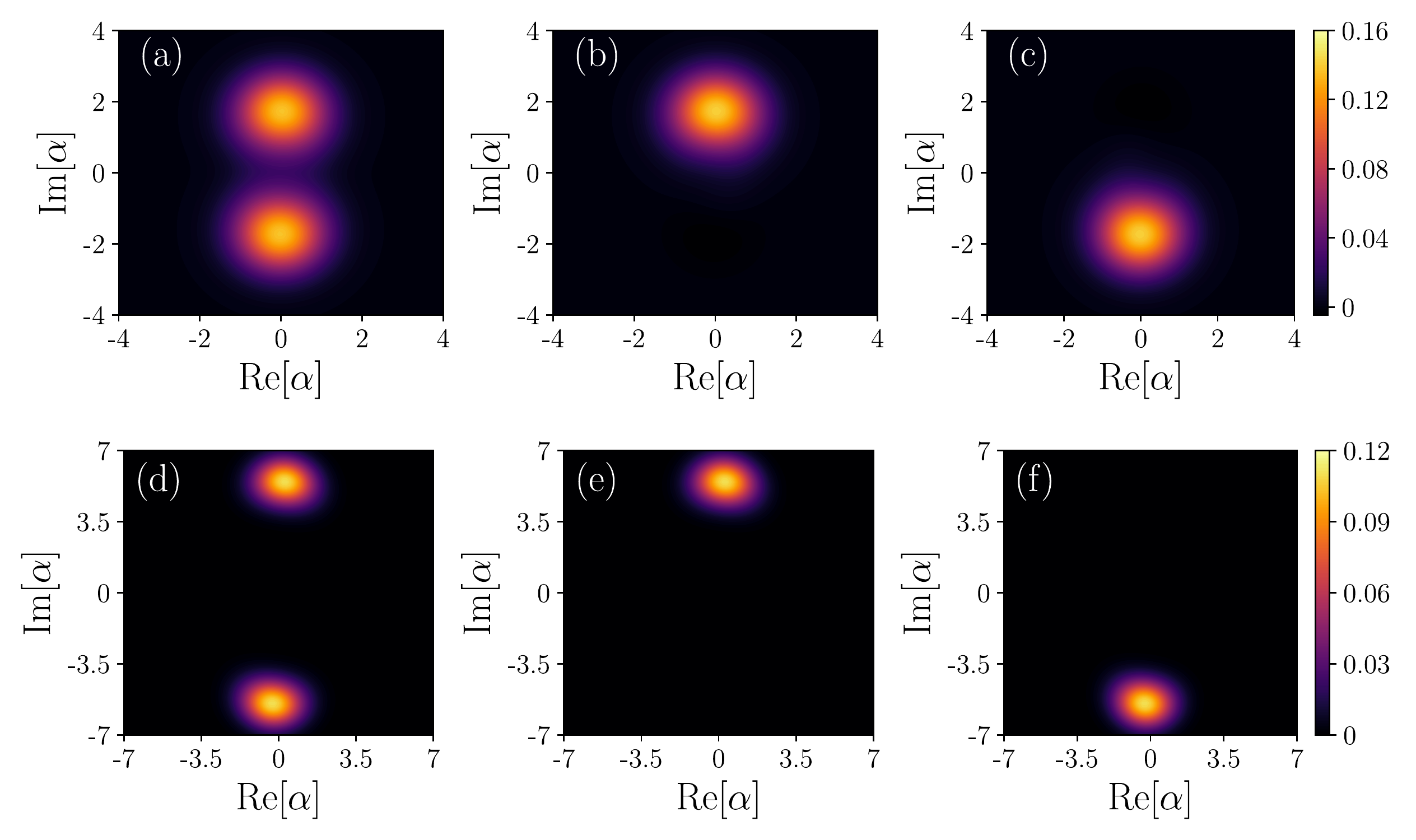}\\
\caption{Colormaps: Wigner distribution $W(\alpha,\alpha^*)$, where $\alpha$ is the amplitude of a coherent state, in different cases: (a) of $\hat{\rho}_{ss}$, (b) of $\hat{\mu}_1/2$, (c) of $\hat{\mu}_2/2$, where the factor $1/2$ is introduced for display purposes. Parameters: $\gamma_2/\gamma_1=3$, $\eta/\gamma_1=2$, $\Delta/\gamma_1=0.1$. In panels (d) to (f) we plot the same as in (a) to (c) but for $\gamma_2/\gamma_1=0.1$, $\eta/\gamma_1=0.2$ and $\Delta/\gamma_1=0.1$.}\label{F2}
\end{figure}

As we show here, the characterization of the EMSs $\hat{\mu}_{1,2}$ provides a first important insight on the relation between the metastable dynamics and entrainment. In fact, we can gain intuition from their visual rendering as provided by their Wigner representation. In Fig. \ref{F2}  we plot the Wigner distribution for $\hat{\rho}_{ss}$, and $\hat{\mu}_{1,2}$. As illustrated in Fig. \ref{F2} (a) and (d) the Wigner distribution of the stationary state displays a manifest bimodal character. Here we consider two disparate parameter values $\gamma_2/\gamma_1=(0.1,3)$ and $\eta/\gamma_1=(0.2,2)$, both satisfying $\Gamma_1/\Gamma_2\ll1$ as can be checked from Fig. \ref{F1}. In fact, we find this bimodality to be present in the whole metastable regime. This can be intuitively understood by considering the Wigner distribution of $\hat{\mu}_{1,2}$, as shown in panels (b), (c) and (e), (f). We can appreciate that each of the lobes corresponds indeed to one of the metastable states.

The bimodality of the stationary state is a reminiscence of classical bistability. However, in stark contrast to the classical case, according to the quantum formalism  these states are metastable, and the effective dynamics of Eq. (\ref{eff_dyn}) tells us that quantum fluctuations eventually drive the system to an even mixture of both. In fact, if we consider an initial condition consisting only of one lobe, i.e. $p_j(0)=1$ and $p_k(0)=0$ with $j\neq k$, we can see  how on times of the order of $\tau_1$ the population of both lobes becomes progressively the same, until the stationary state is finally reached.

In fact, the Wigner distribution of the stationary state is a well known indicator of the emergence of phase-locking between the oscillator and the external forcing \cite{Lee2013,Bruder2014,Bruder2016,Sonar2018}, or between coupled oscillators \cite{Marquardt,Lee2014,Bruder2015}. Indeed, for  the driven QvdP oscillator, a sufficient strong forcing transforms the stationary Wigner distribution from a ring-like shape, indicating lack of a preferred phase, to a localized lobe, indicating the emergence of a preferred phase \cite{Lee2013,Bruder2014}. The latter corresponds to phase-locking in the quantum regime, in which quantum fluctuations play a non-negligible role \cite{Weiss}.  Moving to the case of the squeezed QvdP oscillator, the emergence of phase preference is accompanied by the bimodality of the Wigner distribution \cite{Sonar2018} as displayed in Fig. \ref{F2}. Therefore, each of the EMSs $\hat{\mu}_{1,2}$ can be interpreted as one of the two possible preferred phases to which the QvdP oscillator settles for large enough squeezing strength. It follows that the incoherent process between these two {\it metastable} preferred phases limits phase-locking in the long-time limit, as the stationary state that is reached is an even incoherent mixture of both. The manifestation of this incoherent process limiting phase-locking in different dynamical quantities is explored in more detail in the following section.

Beyond their phase-space representation, the properties of the metastable states can be understood from the fact that they are a linear combination of $\hat{\rho}_{ss}$ and $\hat{\rho}_1$ with $c_{max}=-c_{min}$, and from the symmetries of the Liouvillian. Before proceeding further, notice that  our master equation is invariant under the transformation $\hat{a}\to-\hat{a}$, $\hat{a}^\dagger\to-\hat{a}^\dagger$, i.e. it displays parity-symmetry \cite{Minganti2018,VAlbert}. This means that we can define the unitary transformation $\mathcal{Z}_2\hat{\rho}=e^{i\pi \hat{a}^\dagger \hat{a}}\hat{\rho} e^{-i\pi \hat{a}^\dagger \hat{a}}$, whose action commutes with that of the Liouvillian $[\mathcal{Z}_2,\mathcal{L}]\hat{\rho}=0$. Hence, the eigenmodes are either parity symmetric or parity antisymmetric, which means $\mathcal{Z}_2\hat{\rho}_j=z_j\hat{\rho}_j$ with $z_j=\pm1$, respectively \cite{Minganti2018,VAlbert}. 
An important observation is that the stationary state and the longest-lived eigenmode belong to different symmetry sectors:  
\begin{equation}
\mathcal{Z}_2\hat{\rho}_{ss}=\hat{\rho}_{ss},\quad  \mathcal{Z}_2\hat{\rho}_{1}=-\hat{\rho}_{1}.
\end{equation}
While the stationary state must be parity symmetric \cite{Minganti2018,VAlbert}, the symmetry of $\hat{\rho}_1$ is assessed numerically by realizing that its contribution for parity symmetric observables vanishes identically, i.e. $\text{Tr}[\hat{\rho}_1(\hat{a}^\dagger)^m\hat{a}^n]=0$ for $m+n=\text{even}$, while its contribution for parity antisymmetric ones is generally non-zero, i.e. $\text{Tr}[\hat{\rho}_1(\hat{a}^\dagger)^m\hat{a}^n]\neq0$ for $m+n=\text{odd}$, where $m$ and $n$ are integers. As a consequence of this and of  $c_{max}=-c_{min}$, it follows that  the EMSs do not have a well defined parity-symmetry, and actually they are parity-broken states satisfying:
\begin{equation}
\mathcal{Z}_2\hat{\mu}_{1(2)}=\hat{\mu}_{2(1)}. 
\end{equation}
Thus, both EMSs yield the same value for the expected value of parity symmetric observables, while for parity antisymmetric ones their value solely differs by a phase $e^{i\pi}$.  Then, parity-symmetric observables are insensitive to this metastable dynamics, as for any $\mathcal{P}\hat{\rho}(t)$ [see Eq. (\ref{entrainment_EMSs2})] both $\hat{\mu}_{1,2}$ contribute exactly the same, making irrelevant the particular evolution of $p_{1,2}(t)$ as they sum is one for all $t$. In contrast, notice the particular case of the amplitude\footnote{We use the notation $\langle\hat{o} \rangle_{1(2)}=\text{Tr}[\hat{o}\hat{\mu}_{1(2)}]$, and $\langle\hat{o} \rangle_{ss}=\text{Tr}[\hat{o}\hat{\rho}_{ss}]$.} in which since it is an antisymmetric observable we have $\langle \hat{a}\rangle_1=-\langle \hat{a}\rangle_2$. This is precisely the same phase relation between the two possible classical amplitudes in the synchronized regime, as briefly commented in Sec. \ref{SecModel}. In this sense, it turns out that, as the classical limit is approached, observables computed over the EMSs approach the corresponding classical values for each bistable fixed point \cite{us}. These observations together with the Wigner representations of Fig. \ref{F2}, suggest an intimate connection between the emergence of preferred phases in the quantum regime, and thus synchronization,  and metastability.

\section{Metastable entrained dynamics}

In this section we consider how the characteristic metastable response of the system manifests in the amplitude dynamics and two-time correlations,
in order to characterize entertainment. We find that the metastable regime corresponds to the regime in which the system is entrained by the external forcing. Still, the temporal coherence of this entrained response is ultimately limited by the incoherent process between the two possible preferred phases. Finally, we find that while the long-time amplitude dynamics can be understood from a classical stochastic process, the quantum nature of the fluctuations can manifest in the two-time correlations. 

\subsection{Amplitude dynamics}

\begin{figure}[t!]
\includegraphics[width=0.9\linewidth]{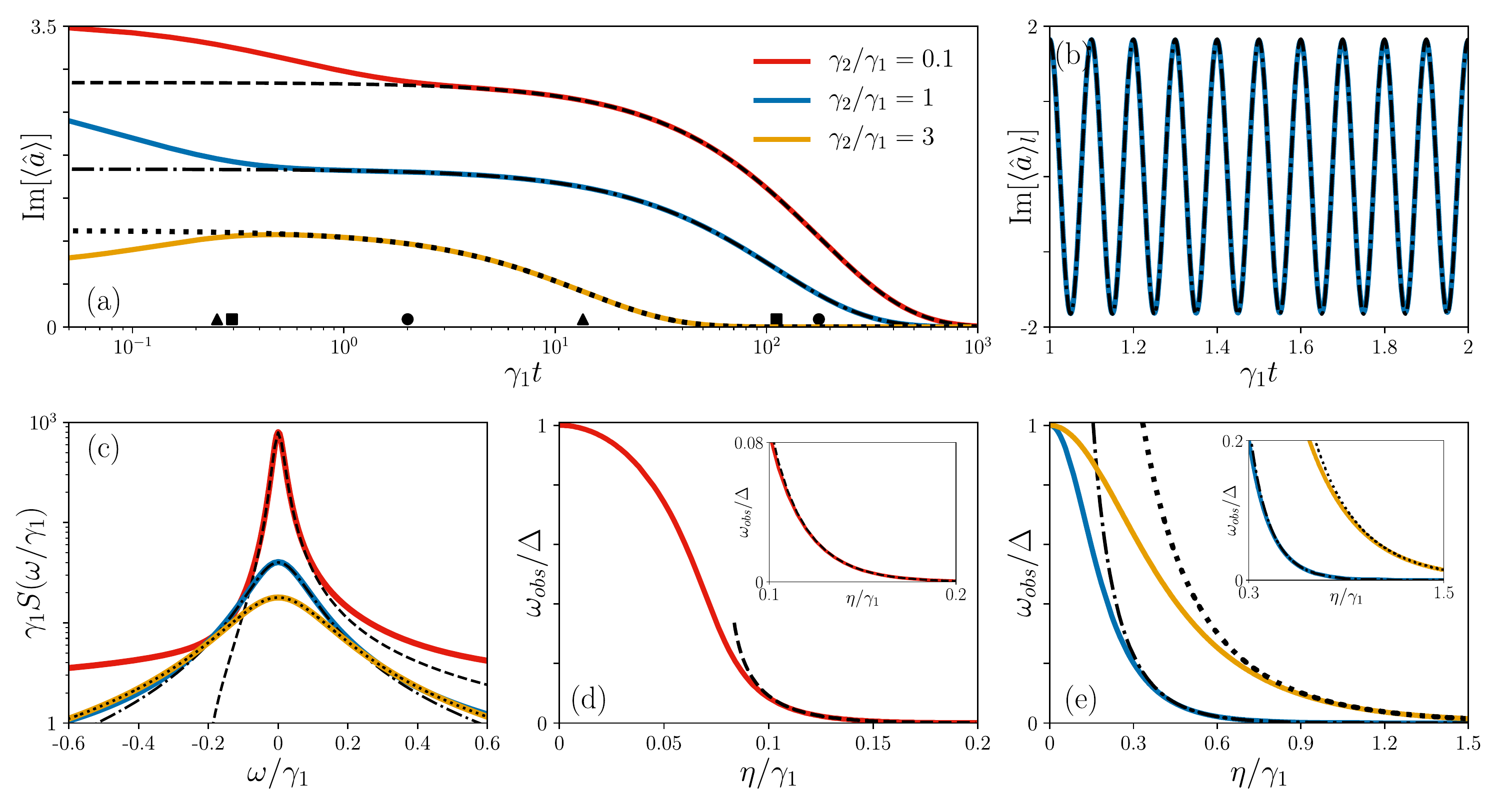}\\
\caption{(a) Imaginary part of $\langle \hat{a}(t) \rangle$ in the rotating frame
for different initial conditions and parameters. Red solid line: $\gamma_2/\gamma_1=0.1$, $\eta/\gamma_1=0.2$, $\hat{\rho}(0)=|1.2\alpha_{+}\rangle\langle 1.2\alpha_{+}|$, where $|x\alpha_{+}\rangle$ is a coherent state of amplitude $x$ times the classical solution $\alpha_{+}$ (\ref{Sec_class}) corresponding to the considered parameters. Blue solid line: $\gamma_2/\gamma_1=1$, $\eta/\gamma_1=1.5$, $\hat{\rho}(0)=|1.5\alpha_{+}\rangle\langle 1.5\alpha_{+}|$. Yellow solid line: $\gamma_2/\gamma_1=3$, $\eta/\gamma_1=2.5$, $\hat{\rho}(0)=|0.5\alpha_{+}\rangle\langle 0.5\alpha_{+}|$. Broken black lines: results using the approximate dynamics of Eq. (\ref{amplitude_dyn}). The markers on the time axis indicate $\tau_1$ (right) and $\tau_2$ (left) for the different parameters: $\gamma_2/\gamma_1=0.1$ (circles), $\gamma_2/\gamma_1=1$ (squares), $\gamma_2/\gamma_1=3$ (triangles). (b) Zoom in of the $\gamma_2/\gamma_1=1$ case in the laboratory frame for $\omega_s/\gamma_1=20\pi$ and $\Delta/\gamma_1=0.1$. In this case the black-broken line corresponds to Eq. (\ref{entrainment}). (c) Emission spectrum considering the exact dynamics (color solid) and the approximate ones (black-broken) for $\gamma_2/\gamma_1=(0.1,1,3)$ and $\eta/\gamma_1=(0.15,0.75,1.875)$ such that $\Gamma_1/\Gamma_2\approx0.05$, with the same color code as in (a). (d),(e) Ratio of the observed frequency and the detuning $\omega_{obs}/\Delta$, considering the exact results (color solid) lines and the approximate ones (black-broken). The lines correspond to $\gamma_2/\gamma_1=(0.1,1,3)$ with the same color code as in (a). The insets show a zoom in of the well-entrained region. In all cases $\Delta/\gamma_1=0.1$.}\label{F3}
\end{figure}

After a short transient of the order of $\tau_2$, the amplitude dynamics for an  arbitrary initial state is restricted to the metastable manifold and is well approximated by
\begin{equation}\label{amplitude_dyn}
\langle \hat{a}(t)\rangle\approx\langle \hat{a} \rangle_1 p_1(t)+\langle \hat{a} \rangle_2 p_2(t),
\end{equation}
where $p_{1,2}(t)$ are the solutions of Eq. (\ref{eff_dyn}).  Notice that the expectation value of the amplitude is defined as $\langle \hat{a}(t)\rangle=\text{Tr}[\hat{a}\hat{\rho}(t)]$, which can be written in terms of the Liouvillian eigenmodes as detailed in \ref{Sec_Liouvillian}. Then, within the approximation of the long-time transient given in Eq. (\ref{MMdynamics}), and using the EMSs decomposition of the metastable manifold, we can obtain the expression given in Eq. (\ref{amplitude_dyn}). The accuracy of Eq. (\ref{amplitude_dyn}) is numerically confirmed in Fig. \ref{F3} (a) for three different values of $\gamma_2/\gamma_1$. In this panel we plot in logarithmic scale the imaginary part of $\langle \hat{a}(t)\rangle$ (the real part would be similar) comparing the exact results of the full model  (color solid lines) with the ones following the reduced EMSs effective long-time dynamics (black broken lines). We see that the initial excited (coherent)  state rapidly relaxes to the metastable manifold  which is accurately described by Eq. (\ref{amplitude_dyn}). Notice that Eq. (\ref{amplitude_dyn}) is the same we would obtain from the two-state stochastic process using the rules of classical stochastics \cite{Gardiner}, which state that the average dynamics of a first moment is the sum of the contributions of each state weighted by $p_{1,2}(t)$ (see also  \ref{Sec_class_twostate}). 

The behavior observed in Fig. \ref{F3} (a) can be fully understood recalling the Liouvillian analysis of Sec.  \ref{sec_meta_Liouvillian}, where we found that whenever  $\Gamma_1/\Gamma_2\ll1$, there is a well defined intermediate time scale $\tau_2\ll t\ll \tau_1$ in which the  dynamics is apparently stable, as the contributions of the modes with $j\geq2$ have already decayed out while the final decay is not yet appreciable as $\Gamma_1 t\ll1$. Thus $\langle \hat{a}(t)\rangle\approx \langle \hat{a} \rangle_1 p_1(0)+\langle \hat{a} \rangle_2 p_2(0)$ on this time scale. This corresponds to the plateaus displayed for intermediate times in Fig. \ref{F3} (a), in between the corresponding markers in the time axis (indicating $\tau_{1,2}$), and which we now show to be intimately related to entrainment. 

The tight connection between metastability and  entrainment can be established going back to the laboratory frame\footnote{We use the notation $\langle\hat{O}(t)\rangle_L$ to designate observables in the laboratory frame. Recall that $\hat{\rho}_L(t)=\hat{U}_t\hat{\rho}(t)\hat{U}^\dagger_t$, from which temporal oscillations at $\omega_s$ or multiples of it can be easily recovered.} where for this intermediate timescale we find  the approximate solution
\begin{equation}\label{entrainment}
\langle \hat{a}(t)\rangle_L\approx\langle\hat{a}\rangle_1 e^{-i\omega_s t}(p_1(0)-p_2(0)),
\end{equation}
According to this, the QvdP oscillator displays an apparently stable perfectly entrained subharmonic response. Provided that $\omega_s\gg\gamma_1$,\footnote{Notice that this condition does not constitute an actual dynamical constraint, at least theoretically, as the parameter $\omega_s$ does not play any role in the rotating frame. Instead, what is physically meaningful is $\Delta=\omega_0-\omega_s$.} the system will  oscillate coherently for many cycles at exactly half of the frequency of the forcing until the effects of the incoherent process described by Eqs. (\ref{eff_dyn}) and (\ref{amplitude_dyn}) start to be evident at times of the order of $\tau_1$. This is exemplified in panel (b), in which we compare the approximate non-decaying subharmonic response of Eq. (\ref{entrainment}) (black-broken line) with the exact result (color solid line) and for intermediate times inside the plateau, finding a very good agreement.

While in a long transient entrainment is then achieved in presence of metastability,  Eq. (\ref{amplitude_dyn}) also illustrates precisely how the incoherent process between the two metastable phases eventually hinders the  1:2 entrained response (or period-doubled dynamics) in the long-time limit. In particular, as the stationary state is a symmetric mixture of both phases, and since they differ by a $\pi$-phase, the expected value of the amplitude eventually decays out on the long timescale given by $\tau_1$ following the dynamics given by Eqs. (\ref{eff_dyn}). Again, this is to be contrasted with the classical case, in which the system settles in one of the bistable locked-phases oscillating subharmonically forever,  due to the absence of fluctuations connecting both phases.

\subsection{Two-time correlations and observed frequency}

We now consider the dynamics of the amplitude two-time correlation in the stationary state, as from this two-time correlation  a well-known indicator of frequency entrainment can be computed. This is the observed frequency \cite{Bruder2014,Bruder2015,Sonar2018,Kato2019}, $\omega_{obs}$, defined as the maximum of the emission or power spectrum\footnote{For models displaying parity symmetry, as in our case, we have that $\langle\hat{a}\rangle_{ss}=0$ and thus the emission spectrum as defined in Eq. (\ref{emissionSpec}) coincides with the fluctuation spectrum or power spectrum as defined by \cite{Kato2019} $S_{inc}(\omega)=\int_{-\infty}^\infty d\tau e^{-i\omega \tau} [\langle\hat{a}^\dagger(\tau)\hat{a}(0)\rangle_{ss}-\langle\hat{a}^\dagger(\tau)\rangle_{ss}\langle\hat{a}\rangle_{ss}]$   (see also \ref{Sec_Liouvillian}). Notice that in some contexts $S_{inc}(\omega)$ is known as the incoherent part of the emission spectrum \cite{Carmichael}.}:
\begin{equation}
\omega_{obs}=\text{argmax}[S(\omega)]
\end{equation}
with
\begin{equation}\label{emissionSpec}
S(\omega)=\int_{-\infty}^\infty d\tau e^{-i\omega \tau}\langle\hat{a}^\dagger (\tau)\hat{a}(0)\rangle_{ss}]
=2\text{Re}\big[\int_0^\infty d\tau e^{-i\omega \tau}\langle\hat{a}^\dagger (\tau)\hat{a}(0)\rangle_{ss}\big]
\end{equation}
where the subscript 'ss' denote that this correlation is calculated in the stationary state  in which two-time correlations only depend on the difference between the time arguments, i.e.  $\langle\hat{a}^\dagger (-\tau)\hat{a}(0)\rangle_{ss}=\langle\hat{a}^\dagger (0)\hat{a}(\tau)\rangle_{ss}$, while we have also used the fact that $\langle\hat{a}^\dagger (0)\hat{a}(\tau)\rangle_{ss}=[\langle\hat{a}^\dagger (\tau)\hat{a}(0)\rangle_{ss}]^*$ (see \ref{Sec_Liouvillian}).

Perfect (subharmonic) entrainment arises when the system oscillates at (half of) the driving frequency. As Eq. (\ref{emissionSpec}) is
in the rotating frame, the observed frequency will be
zero when the system is well entrained, while it
is known to be close to the intrinsic detuning, $\Delta$, when there is no synchronization  \cite{Bruder2014,Sonar2018}. In fact, in the metastable regime the emission spectrum is dominated by the contribution of the longest-lived eigenmode. Intuition about this can be gained from the eigendecomposition of the relevant two-time correlation,  Eq. (\ref{TwoTime1}) in  \ref{Sec_Liouvillian}. Since in the metastable regime we have that $\Gamma_1/\Gamma_{j\geq2}\ll1$, the resonance associated with this mode should stand out in the spectrum. Indeed, we show in Fig. \ref{F3} (c) how the contribution of this mode fits accurately the main peak of the spectrum in this regime.  Furthermore this panel also illustrates how $\Gamma_1$ diminishes  when decreasing  $\gamma_2/\gamma_1$: in the three cases amplification/damping rates and squeezing are varied but maintaining $\Gamma_1/\Gamma_2\approx0.05$, and we can see how diminishing $\gamma_2/\gamma_1$ the resonances become significantly sharper, a signature of the vanishing of $\Gamma_1$ in the classical limit \cite{us}. In fact, the decay rate of the dominant mode takes the values $\Gamma_1/\gamma_1\approx(0.017,0.087,0.152)$ for the cases $\gamma_2/\gamma_1=(0.1,1,3)$, respectively. We further notice that in Fig. \ref{F3} (c) we are varying the squeezing strength and the non-linear damping at the same time, both parameters affecting the width of the peaks. In particular, as we have seen, the width of the main peak is accurately captured by $\Gamma_1$, which diminishes with increasing squeezing strength above the EP, and the rest of parameters fixed [e.g. see Fig. \ref{F1} (d)]; as for $\Gamma_1$, it increases with the non-linear damping strength and the rest of parameters fixed (not shown here).

While we recall that the large-$\tau$ dynamics of two-time correlations can be written in terms of Eq. (\ref{eff_dyn}) \cite{Garrahan1,Garrahan2}, we find that it is more illustrative to write it in terms of the (fully equivalent) Eq. (\ref{MMdynamics}). In particular from   Eqs. (\ref{TwoTime1}) and (\ref{MMdynamics}) it follows that for $\Gamma_1/\Gamma_2\ll1$:
\begin{equation}\label{eff_TwoTime}
 \langle \hat{a}^\dagger(\tau)\hat{a}(0)  \rangle_{ss}\approx\text{Tr}[\hat{\sigma}_1 \hat{a}\hat{\rho}_{ss}]\text{Tr}[\hat{a}^\dagger\hat{\rho}_1] e^{-\Gamma_1\tau}.
\end{equation}
We highlight the factor $\text{Tr}[\hat{\sigma}_1 \hat{a}\hat{\rho}_{ss}]$ in this equation stemming from the quantumness of the model. Indeed a two-time correlation (and thus the emission spectrum) follows from an initial perturbation of the stationary state (here $\hat{a}\hat{\rho}_{ss}$) \cite{Carmichael}, and  can be interpreted as the unavoidable disturbance of a measurement process. Generally, this factor makes this two-time correlation different from the corresponding  one of a classical two-state stochastic process \cite{Gardiner}, which is found to be $C(\tau)=|\langle\hat{a}\rangle_1|^2 e^{-\Gamma_1\tau}$ (see  \ref{Sec_class_twostate}). Since generally  $\text{Tr}[\hat{\sigma}_1 \hat{a}\hat{\rho}_{ss}]\neq \langle \hat{a}\rangle_1$, we find that $\langle \hat{a}^\dagger(\tau)\hat{a}(0)  \rangle_{ss}\neq C(\tau)$. This manifests in the fact that the Fourier transform of $C(\tau)$ is centered at the origin for $\eta\geq\eta_{EP}$ while this is not the case for the quantum case, as we shall see in Fig. \ref{F3} (d) and (e). Thus, in contrast to the case of the long-time amplitude dynamics Eq. (\ref{amplitude_dyn}),  multi-time correlations do not follow straightforwardly from the corresponding classical law of a two-state stochastic process. In these two-time correlations, the quantum nature of the fluctuations and of the degrees of freedom becomes manifest. 

We now proceed to analyze the behavior of the observed frequency. In Figs. \ref{F3} (d) and (e) we exemplify for different parameter values how as the squeezing strength $\eta$ increases the system becomes entrained, i.e. $\omega_{obs}/\Delta$ goes from one to zero. Notice how the transition to entrainment is sharper the smaller is $\gamma_2/\gamma_1$ (large number of excitations). For $\eta>\eta_{EP}$ we have plotted in black-broken lines the observed frequency calculated from the effective long-time dynamics, i.e. Eq. (\ref{eff_TwoTime}). Here, we can observe that the asymptotic decay of $\omega_{obs}$ towards zero is very well captured by this approximate calculation. This can be further appreciated in the insets of these two panels, in which the well-entrained region, that of $\omega_{obs}$ close to zero, is shown in more detail. This constitutes a further confirmation of what we have been discussing previously: the fact that the entrained dynamics is characterized by the metastable response of the state of the system. Thus the signatures of entrainment in the power spectrum are captured by the effective long-time dynamics between the metastable phases, that follows from the opening of a spectral gap.

\section{Discussion and conclusions}\label{Sec_Conclusions}

In this work we have established the connection between the quantum entrainment in the squeezed QvdP oscillator and quantum metastability.  We have reported that squeezing enables the opening of a spectral gap in the Liouvillian, which leads to a huge separation of timescales in the dynamics: after a short transient of time the system settles into the so-called metastable manifold in a mixture of two metastable states  that depends on the initial condition. It then follows an incoherent process between the two parity-broken metastable preferred phases of the entrained oscillator. Indeed, the oscillation frequency settles to the value of half of  the forcing. Still,  quantum entrainment is ultimately limited by this incoherent process,  as on a long time scale  the temporal coherence of the subharmonic response eventually decays out. In this sense, future investigations could address the  question of how the reported metastable entrained dynamics manifests at the {\it quantum trajectory level}, analyzing several possible stochastic unravellings of our master equation \cite{WisemanMilburn}, while considering the dynamics of different observables and comparing with the fixed point attractor scenario of the driven QvdP oscillator,  or that of coupled QvdP oscillators \cite{Manzano2}.

The distinctive features of the analyzed quantum entrained dynamics stem from the distinct classical attractors introduced by the squeezed forcing. Hence, classical phase bistability becomes phase metastability in the presence of quantum fluctuations, as reflected by the dominant fluctuation mode in the system. This more complex scenario is to be contrasted with quantum entrainment in the driven QvdP oscillator (in the absence of squeezing) \cite{Lee2013,Bruder2014,Weiss}. There the dominant fluctuation modes describe the dynamics around the unique fixed point attractor of the entrained regime \cite{Weiss}, which can indeed be analyzed by linearizing the master equation around this stable point \cite{Weiss}. This qualitative difference is behind the reported specific features of the {\it power} or {\it incoherent} spectrum (i.e. the part of the emission spectrum corresponding to the fluctuations) \cite{Sonar2018,Weiss}. In fact, in Ref. \cite{Weiss} this fluctuations dynamics is analyzed in detail as well as its signatures in the power spectrum. Fluctuations are shown to display an overdamped regime and an underdamped one, which translate in the power spectrum displaying either a broad peak centered at the driving frequency or displaying sidebands around this frequency, respectively \cite{Weiss}. Moreover, in the overdamped regime the width of the Lorentzian-like peak is reported to increase with the driving strength  \cite{Sonar2018}, which means that the fluctuations or perturbations around the fixed point attractor decay faster as the system becomes better entrained. This can be interpreted as this attractor becoming more stable and less susceptible to the effects of fluctuations as the forcing strength is increased, i.e. to the coherent part of the dynamics becoming enhanced as the system becomes entrained, while the incoherent spectrum flattens. On the other hand, in the case of the squeezed QvdP oscillator, the width of the main Lorentzian peak decreases significantly as the squeezing strength is increased with the other parameters fixed as reported here [see the behavior of $\Gamma_1$ in Fig. \ref{F1} (d)] or in \cite{Sonar2018}. As our analysis reveals, this contrasting behavior of the two cases is due to the completely different physical origin of this peak, rooted in the fluctuation dynamics {\it between} the two fixed point attractors of the entrained regime. Thus, the decreasing of this linewidth is to be interpreted as the jumps between the two preferred phases becoming suppressed (or less frequent), which also leads to an enhanced temporal coherence for the entrained subharmonic response. Therefore, we conclude that a careful assessment of the fluctuation dynamics is crucial to understand the synchronized response of quantum systems and its key signatures in the power spectrum.

Precisely, the characteristic fluctuation dynamics reported here transcends the particular context of the squeezed QvdP oscillator. As discussed, metastability appears in many different driven-dissipative quantum systems \cite{Garrahan1}. A particularly relevant example for us is the case of DPTs with spontaneous parity-symmetry breaking \cite{Minganti2018}. This is indeed what happens in this system in the infinite-excitation limit where the classical attractors emerge, and the parity-broken metastable preferred phases acquire a divergent lifetime \cite{us}. Other examples of metastable dynamics associated with parity-breaking transitions have been reported in Refs. \cite{Minganti2018,Minganti2016,Bartolo2016b}. Moreover, such long-time scales associated with fluctuations between multiple possible phases have also been reported for quantum systems of parametric oscillators \cite{Dykman}, period-tripling oscillators \cite{Kubala}, and optomechanical oscillators \cite{Weiss2}. Thus, slow-relaxation time scales seem a characteristic feature of multistable dynamical systems in the quantum regime, and a seemingly metastable manifold and dynamics could be expected for these examples as well as in further synchronization scenarios with multiple preferred phases. In this sense, it would also be interesting to investigate whether the metastable entrained dynamics reported here can be seen as a form  of  quantum activation process \cite{Marthaler2006,Dykman2007} emerging in a far-from-equilibrium scenario, in which {\it non-linear} dissipation plays a fundamental role in shaping the properties of the metastable states.

Finally, the presence of a long-lived collective excitation has been found to be behind the emergence of {\it spontaneous} synchronization in a variety of open quantum systems. This is the case of transient synchronization  recently reviewed in Ref. \cite{SyncRev2}. In fact,  such an excitation can display a lifetime much longer than the rest, with the extreme possibility of attaining no-decay. The latter corresponds to  stationary synchronization, being associated with highly symmetric situations in the presence of noiseless subsystems and decoherence free subspaces \cite{Giorgi1,Manzano_PRA,Manzano_SciReps,Bellomo,Cabot_NPJ}, as well as strong dynamical symmetries \cite{Dieter2}. Transient synchronization and the reported metastable entrained dynamics have in common that the emergence of synchronization  is associated with a significant timescale separation in the dynamics, i.e. the opening of a spectral gap, that leads to a long-lived synchronized response practically independent of the initial conditions \cite{SyncRev2}. In a bigger picture, these examples also illustrate the fact that, while synchronization in quantum systems is possible, the temporal coherence of such a dynamical response is often limited by the presence of quantum fluctuations inherent in open quantum systems, generally leading to  a finite time-window for the observation of the synchronized dynamics, in stark contrast to the case of classical noiseless dynamical systems \cite{Pikovsky,Belanov}. 

\section*{Acknowledgements}

We acknowledge interesting comments from Rosario Fazio and the use of the python package QuTip \cite{Qutip1,Qutip2} in order to obtain some of the numerical results. We acknowledge the Spanish State Research Agency, through the   Severo Ochoa and María de Maeztu Program   for   Centers   and   Units   of   Excellence   in R\&D  (MDM-2017-0711)  and  through  the  QUARESC project (PID2019-109094GB-C21and-C22/AEI/10.13039/501100011033); We  also    acknowledge funding   by   CAIB   through   the   QUAREC   project (PRD2018/47).  AC  is funded by the CAIB PhD program. GLG is funded by the Spanish Ministerio de Educación y Formación Profesional/Ministerio Universidades and co-funded by the University of the Balearic  Islands  through  the  Beatriz  Galindo  program (BG20/00085).

\appendix

\section{Liouvillian approach to the dynamics}\label{Sec_Liouvillian}

As commented in the main text, the Liouvillian superoperator can be diagonalized obtaining a set of eigenvalues $\lambda_j$, and the corresponding right and left eigenmatrices $\hat{\rho}_j$ and  $\hat{\sigma}^\dagger_j$, respectively.  These eigenvalues are non-positive and either real or appear in complex conjugate pairs. In the former case, the corresponding eigenmatrices can be defined Hermitian \cite{Minganti2018}. Notice that except for spectral singularities \cite{Minganti2019}, these eigenmatrices form a biorthogonal basis that can be normalized by means of the Hilbert-Schmidt product: $\text{Tr}[\hat{\sigma}_j^\dagger \hat{\rho}_k]=\delta_{jk}$. Moreover, in case the real part of the eigenvalues is negative, the corresponding eigenmodes are traceless, and hence they are not physical states \cite{Minganti2018}. In this case we can choose whether to rescale $\hat{\sigma}^\dagger_{j}$ or $\hat{\rho}_{j}$ freely, as long as they satisfy the normalization condition. The stationary state is obtained from the properly normalized right eigenmatrix with zero eigenvalue, i.e. $\hat{\rho}_{ss}=\hat{\rho}_{0}/\text{Tr}[\hat{\rho}_0]$ while the left one is the identity $\hat{\sigma}_0= {I}$.  

Since except for spectral singular points these eigenmatrices form a basis of the system Hilbert space, they can be used to decompose the state of the system at any time as done in Eq. (\ref{general_dynamics}). By taking the trace over a given operator we can obtain its dynamics. In particular for the amplitude dynamics we obtain: 
\begin{equation}\label{amp_decay}
\langle \hat{a}(t)\rangle=\sum_{j\geq1}\text{Tr}[\hat{\sigma}_j^\dagger\hat{\rho}(0)]\text{Tr}[\hat{a}\hat{\rho}_j]e^{\lambda_j t}.
\end{equation}
 
Here we have used the fact that the Liouvillian  displays a parity symmetry, and thus the eigenmodes are either parity symmetric or parity antisymmetric, and the stationary state is parity symmetric.  In fact, it follows that for operators with a given symmetry, only eigenmodes with the same symmetry contribute to their dynamics. For the particular case of the amplitude only  parity antisymmetric eigenmodes contribute to the sum, as $\text{Tr}[\hat{a}\hat{\rho}_j]=0$ is identically zero if $\mathcal{Z}_2\hat{\rho}_j=\hat{\rho}_j$. Notice that this is also the case for the amplitude two-time correlations calculated in the stationary state. But first, let us recall how to compute these two-time correlations from the master equation \cite{Carmichael}:
\begin{equation}\label{twotime_1}
\langle \hat{a}^\dagger(t_1)\hat{a}(t_2)\rangle=\text{Tr}[\hat{a}^\dagger e^{\mathcal{L}(t_1-t_2)}\big(\hat{a}\hat{\rho}(t_2)\big)],\text{ if } t_1\geq t_2,
\end{equation}
and
\begin{equation}\label{twotime_2}
\langle \hat{a}^\dagger(t_1)\hat{a}(t_2)\rangle=\text{Tr}[\hat{a} e^{\mathcal{L}(t_2-t_1)}\big( \hat{\rho}(t_1)\hat{a}^\dagger\big)], \text{ if } t_2\geq t_1,
\end{equation}
where the superoperator $e^{\mathcal{L}|t_1-t_2|}$ applies to everything into the  big rounded parenthesis, and it stands for time-evolving for a period of time $|t_1-t_2|$ and according to $\mathcal{L}$ the 'initial condition' inside the big rounded parenthesis. Following \cite{Carmichael}, these correlations can also be described using the variables $\tau\geq0$ and $t=\text{min}(t_1,t_2)$, from which we obtain:
\begin{equation}
\langle \hat{a}^\dagger(t+\tau)\hat{a}(t)\rangle=\text{Tr}[\hat{a}^\dagger e^{\mathcal{L}\tau}\big(\hat{a}\hat{\rho}(t)\big)], \text{ for }\tau\geq0,
\end{equation}
and an analogous expression for the case of (\ref{twotime_2}). The stationary two-time correlations studied in the main text are then defined as 
\begin{equation}
\langle \hat{a}^\dagger(\tau)\hat{a}(0)\rangle_{ss}=\lim_{t\to\infty}\langle \hat{a}^\dagger(t+\tau)\hat{a}(t)\rangle=\text{Tr}[\hat{a}^\dagger e^{\mathcal{L}\tau}(\hat{a}\hat{\rho}_{ss})],
\end{equation}
and analogously for the other correlation. In terms of the Liouvillian eigenmodes we can write:
\begin{equation}\label{TwoTime1}
\langle \hat{a}^\dagger(\tau)\hat{a}(0)\rangle_{ss}=\sum_{j\geq1}\text{Tr}[\hat{\sigma}_j^\dagger \hat{a}\hat{\rho}_{ss}]\text{Tr}[\hat{a}^\dagger\hat{\rho}_j]e^{\lambda_j\tau},
\end{equation}
in which we find again that the term $j=0$ does not contribute as $\langle \hat{a}\rangle_{ss}=0$ due to parity symmetry. Otherwise, in the absence of parity
symmetry there can be a non-zero contribution from $j = 0$, i.e. an additional
term that reads $\langle\hat{a}^\dagger\rangle\langle \hat{a}\rangle_{ss}$. Since $\lambda_0=0$, this term is time-independent and yields a Dirac delta when Fourier transformed. This contribution is known in some contexts as the coherent part of the emission spectrum \cite{Carmichael}.

\section{Classical subharmonic entrainment}\label{Sec_class}

We briefly overview the dynamical regimes that this model displays in the classical limit. In this limit we define the complex amplitude $\alpha=\langle \hat{a} \rangle$, whose equation of motion is obtained from that of $\langle \hat{a}\rangle$ facotrizing higher-order moments:
\begin{equation}\label{class_eq}
\partial_t \alpha=-i\Delta\alpha+\frac{\gamma_1}{2}\alpha-\gamma_2|\alpha|^2\alpha-2\eta\alpha^*. 
\end{equation}

This equation displays a single bifurcation at the classical critical point $\eta_c=|\Delta|/2$. For $\eta<\eta_c$, the stable attractor is a limit-cycle of fundamental frequency $\Omega=\Delta\sqrt{1-(2\eta/\Delta)^2}$. While for $\eta>\eta_c$ the stable attractors are two bistable fixed points that only differ by a complex phase $e^{i\pi}$; their amplitudes are given by $\alpha_\pm=\pm Re^{i\phi}$ with $R=\sqrt{\frac{\gamma_1}{2\gamma_2}+\frac{1}{\gamma_2}\sqrt{4\eta^2-\Delta^2}}$, while the phase is $2\phi=\pi-\sin^{-1}[|\Delta|/(2\eta)]$ \cite{Sonar2018,Kato2021}. Back into the laboratory frame, the limit-cycle regime  corresponds to the lack of entrainment since $\Omega$ and $2\omega_s$ are not generally commensurate. On the other hand, the bistable solutions become $\alpha_\pm(t)=\pm Re^{i\phi-i\omega_s t}$, and thus this regime corresponds to subharmonic entrainment with two possible locked phases \cite{Pikovsky,Belanov}.

\section{Metastable dynamics}\label{Sec_meta_states}

As we have explained in the main text, the long-time dynamics of the system is restircted to the so-called metastable manifold \cite{Garrahan1,Garrahan2}. Notice that since in the whole region in which $\Gamma_1/\Gamma_2\neq1$, $\lambda_1$ is real, we find that $\hat{\sigma}_1$ and $\hat{\rho}_1$ are Hermitian. This property enables to parametrize the metastable manifold in terms of the extreme metastable states (EMSs) defined in Eq. (\ref{EMSs}), as well as the projectors on these states, defined by:
\begin{equation}
\hat{P}_1=\frac{1}{\Delta c}(\hat{\sigma}_1 -c_{min} {I}),\quad \hat{P}_2=\frac{1}{\Delta c}(-\hat{\sigma}_1+c_{max} {I}),
\end{equation}
with $\Delta c=c_{max}-c_{min}$, while $c_{max}$ and $c_{min}$ are the maximum and minimum eigenvalues of $\hat{\sigma}_1$\footnote{Notice that from orthogonality $\text{Tr}[\hat{\sigma}_1\hat{\rho}_{ss}]=0$ it follows that $c_{min}\leq0$, while since $\hat{\sigma}_1$ is Hermitian, it follows that all its eigenvalues are real.}. As commented, numerical analysis reveals that $c_{min}=-c_{max}$ for the considered regime, i.e. when $\Gamma_1/\Gamma_2<1$. 

The projection of the initial state onto the metastable manifold can be decomposed as the projection onto the EMSs: $\mathcal{P}\hat{\rho}(0)=\text{Tr}[\hat{P}_1\hat{\rho}(0)]\hat{\mu}_1+\text{Tr}[\hat{P}_2\hat{\rho}(0)]\hat{\mu}_2$. Importantly, we have that by construction $\hat{P}_{1,2}\ge 0$, $\hat{P}_{1}+\hat{P}_{2}= {I}$ and $\text{Tr}[\hat{P}_i\hat{\mu}_j]=\delta_{ij}$.   Moreover, we can define the expectation values  $p_{1(2)}(t)=\text{Tr}[\hat{P}_{1(2)}\hat{\rho}(t)]$. It follows that $p_{1,2}(t)$ are positive and they sum to one. Hence $p_{1,2}(t)$ can be interpreted as probabilities as stated in the main text. Notice that this is only the case  for the EMSs, as they form the only basis in which $\hat{P}_{1,2}$ are positive in all the metastable manifold \cite{Garrahan1}. The long-time dynamics can be recasted in this basis leading to Eq. (\ref{eff_dyn}), which in virtude of the properties of $p_{1,2}(t)$ can be interpreted as a stochastic process \cite{Garrahan1}.  Then the solution to this equation is: 
\begin{equation}\label{eff_sol}
p_{1(2)}(t)=\frac{p^0_{1(2)}}{2}(1+e^{-\Gamma_1t})+\frac{p^0_{2(1)}}{2}(1-e^{-\Gamma_1t}),
\end{equation}
where the initial conditions are given by the projection of the initial state onto the metastable manifold in terms of the EMSs: $p^0_{1(2)}=\text{Tr}[\hat{P}_{1(2)}\hat{\rho}(0)]$. 

\begin{figure}[t!]
\includegraphics[width=0.75\linewidth]{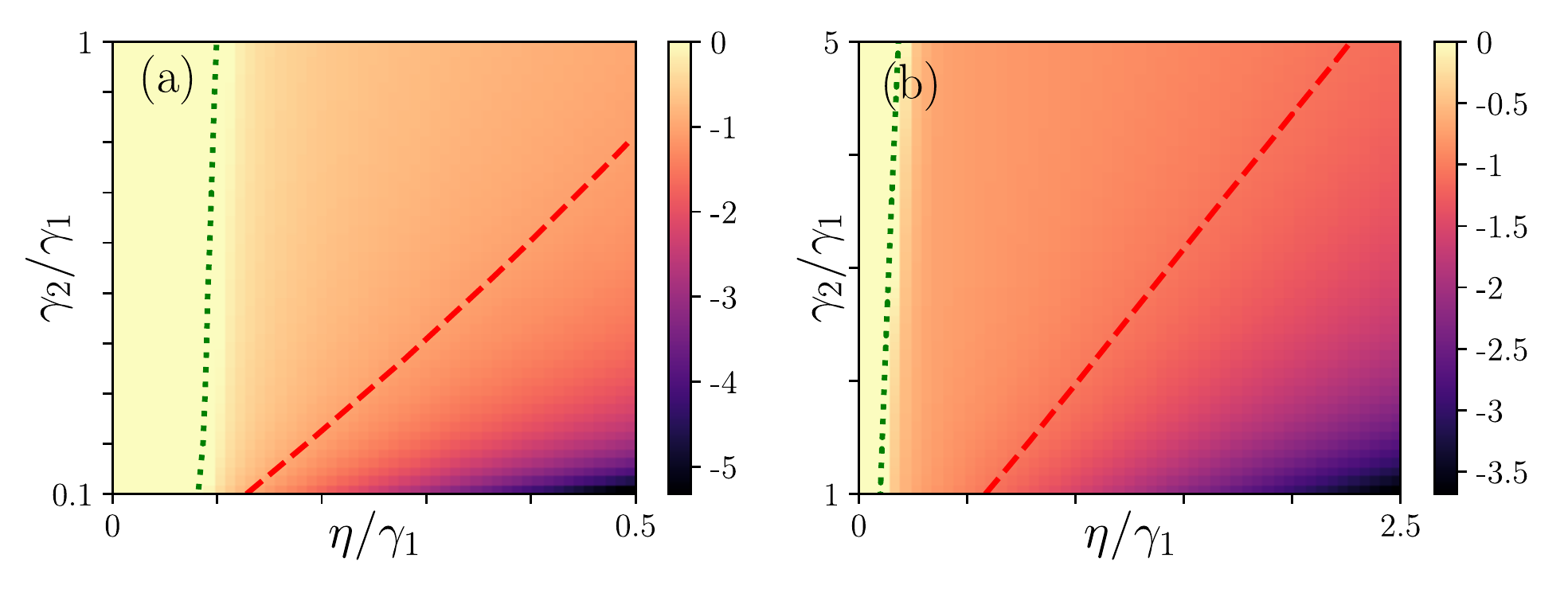}\\
\caption{Colormaps: $\text{log}_{10}[D(\hat{\mu}_1,\hat{\mu}_1')]$ varying $\gamma_2/\gamma_1$ and $\eta/\gamma_1$. Panel (a) covers in detail the small values of $\gamma_2/\gamma_1$ and $\eta/\gamma_1$, while panel (b) displays a wider range of larger values. The regions in which $\Gamma_1=\Gamma_2$, i.e. for $\eta\leq\eta_{EP}$ have been colored in palid yellow. Green-dotted lines: contour indicating the EP for different values of $\gamma_2/\gamma_1$. Red-dashed lines: contours inticating $\Gamma_1/\Gamma_2=0.1$.}\label{F4}
\end{figure}

Finally we compare the exact expressions for the EMSs, i.e. $\hat{\mu}_{1}=\hat{\rho}_{ss}+c_{max}\hat{\rho}_1$ with the approximate ones, that we distinguish in this appendix with a prime, i.e. $\hat{\mu}'_{1}=\hat{\rho}_{ss}+\hat{\rho}_1$. We do so by computing the trace distance between both, defined as $D(\hat{\mu}_1,\hat{\mu}_1')=\text{Tr}[\sqrt{(\hat{\mu}_1-\hat{\mu}_1')^\dagger(\hat{\mu}_1-\hat{\mu}_1')}]/2$, see Fig. \ref{F4}. Notice that in this figure we have assigned the value one to the region below the EP in which $\Gamma_1=\Gamma_2$, which we have delimited by a green-dotted line in order to highlight that it is not part of the analysis. Moreover, we have used a red-dashed line to indicate the contour $\Gamma_1/\Gamma_2=0.1$. Comparing Fig. \ref{F4} (a), (b) with Fig. \ref{F1} (a), (b), we can see how the trace distance becomes vanishingly small as $\Gamma_1/\Gamma_2$ becomes small. Indeed both quantities seem to be very well correlated, as can be checked from these figures. This means that in the regions in which there is a huge separation of timescales, and hence it is meaningful to define the effective dynamics (\ref{eff_dyn}), we have that $\hat{\mu}_1\approx\hat{\mu}_1'$ to a very good approximation. The same result is found for $\hat{\mu}_2$ and $\hat{\mu}_2'$ (not shown). Then, as commented in the main text, in the metastable regime we can use the more convenient expressions for the EMSs and  also for the projectors: $\hat{\mu}_{1,2}'$ and $\hat{P}'_{1(2)}\approx[ {I}+(-)\hat{\sigma}_1]/2$.

\section{Classical two-state stochastic process}\label{Sec_class_twostate}

In this appendix we consider Eq. (\ref{eff_dyn}) as a fully classical two-state stochastic process and we compute the amplitude dynamics and two-time correlations according to the classical rules \cite{Gardiner}. In particular, we consider that state 1 is characterized by the complex amplitude $\alpha_1=\langle\hat{a}\rangle_1$, while state 2 by the complex amplitude $\alpha_2=-\langle\hat{a}\rangle_1$. The solution for the classical two-state process is again given by Eq. (\ref{eff_sol}), while the difference with the effective long-time quantum dynamics resides, in principle, in the recipe to compute statistical quantities. Firstly, we have that the classical averaged amplitude is given by \cite{Gardiner}:
\begin{equation}
\overline{\alpha(t)}=\sum_{j=1}^2 \alpha_j p_j(t), 
\end{equation}
where the overbar indicates a classical average over the stochastic process. As commented in the main text this expression coincides with Eq. (\ref{amplitude_dyn}). Notice that the stationary value for the amplitude is zero, i.e. $\overline{\alpha_{ss}}=0$, as expected. Secondly, the classical recipe for the two-time amplitude correlations in the stationary state is given by \cite{Gardiner}:
\begin{equation}
C(\tau)=\lim_{t\to\infty}\overline{\alpha^*(t+\tau)\alpha(t)}=\sum_{i=1}^2 \sum_{j=1}^2 \alpha_i^*\alpha_j \tilde{p}_{ij}(\tau)p_j(t\to \infty), 
\end{equation}
where we have defined the tilded probabilities $\tilde{p}_{ij}(\tau)$ as the solutions $p_i(\tau)$ given in Eq. (\ref{eff_sol}) with the special initial condition $p_{j}^{(0)}=1$. From this expression we recover the result quoted in the main text, that is $C(\tau)=|\alpha_1|^2e^{-\Gamma_1\tau}$.


\normalem
\bibliographystyle{revtex4-1}
\bibliography{references}
\end{document}